\documentclass[%
twocolumn,amsmath,amssymb,superscriptaddress,preprintnumbers,prl
]{revtex4-1}

\usepackage[dvipdfmx]{graphicx}
\usepackage{multirow}
\usepackage{bm}

\begin{document}
\title{Organic Molecular Dynamics and Charge-carrier Lifetime in Lead Iodide Perovskite MAPbI$_3$}

\author{A. Koda}
\author{H. Okabe}
\author{M. Hiraishi}
\affiliation{ 
Muon Science Laboratory, 
Institute of Materials and Structural Science, 
High Energy Accelerator Research Organization (KEK), 
Tokai, Naka, Ibaraki 319-1106, Japan
}%
\author{R. Kadono}  
\affiliation{ 
Muon Science Laboratory, Institute of Materials and Structural Science, 
High Energy Accelerator Research Organization (KEK), 
Tokai, Naka, Ibaraki 319-1106, Japan
}%
\affiliation{Department of Materials Structure Science, The Graduate University for Advanced Studies, Japan}

\author{K. A. Dagnall}
\author{J. J. Choi}
\affiliation{ 
Department of Chemical Engineering, University of Virginia, Charlottesville, Virginia 22904, USA
}%

\author{S.-H. Lee}
\affiliation{ 
Department of Physics, University of Virginia, Charlottesville, Virginia 22904, USA
}%

\preprint{www.pnas.org/cgi/doi/10.1073/pnas.2115812119}

\begin{abstract}
The long charge carrier lifetime of the hybrid organic-inorganic perovskites (HOIPs) is the key for their remarkable performance as a solar cell material. The microscopic mechanism for the long lifetime is still in debate. Here, by using muon spin relaxation technique that probes the fluctuation of local magnetic fields, we show that the muon depolarization rate ($\Delta$) of a prototype HOIP methylammonium lead iodide (MAPbI$_3$) shows  a sharp decrease with increasing temperature in two steps above 120 K and 190 K across the structural transition from orthorhombic to tetragonal structure at 162 K.  Our analysis shows that the reduction of $\Delta$ is quantitatively in agreement with the expected behavior due to the rapid development of MA jumping rotation around the $C_3$ and $C_4$ symmetry axes.  Our results provide direct evidence for the intimate relation between the rotation of the electric dipoles of MA molecules and the charge carrier lifetime in HOIPs.
\end{abstract} 



\maketitle

Hybrid organic-inorganic perovskites (HOIPs) have been attracting enormous research interest as one of the most promising materials for the next-generation solar cells that combine high efficiency and low cost  \cite{Kojima:09}. The power conversion efficiency (PCE) of HOIP-based solar cells has reached above 25\% \cite{Kim:20}, which is comparable to that of silicon solar cells. In view of the social implementation, the merit is emphasized that HOIP solar cells can be manufactured using simple solution processing methods with drastically low costs compared with the current commercial solar cell technologies. Meanwhile, the most promising HOIP family including the prototype methylammonium lead iodide (MAPbI$_3$, where MA denotes CH$_3$NH$_3^+$) has been shown to be chemically unstable \cite{Yang:17}, and the attempts to replace Pb with less toxic elements have had limited success in achieving high PCE \cite{Ke:17,Shao:18}. Moreover, the microscopic mechanism of the high PCE in the MAPbI$_3$ family is still elusive, despite years of extensive research on their basic properties.

One of the most important properties relevant to the high photovoltaic performance of HOIPs is their long carrier lifetimes ($\gtrsim1$ $\mu$s) observed in thin films, which translates to large carrier diffusion lengths despite their modest charge mobilities \cite{Brenner:16}. Several microscopic mechanisms behind the unusually long carrier lifetime have been proposed, such as the formation of ferroelectric domains \cite{Frost:14,Kutes:14,Strelcov:17}, the Rashba effect \cite{Zheng:15,Etienne:16}, the photon recycling \cite{Yamada:17}, and the formation of large polarons \cite{Chen:16,Zhu:16}. When the HOIPs are replaced with all inorganic perovskites in the photovoltaic architecture, the device can still function as a solar cell. This indicates that the photons excite electrons and holes out of the inorganic metal halide atoms, which is consistent with the density functional theory (DFT) calculations that the corner interstitial cations, whether organic or inorganic, do not directly contribute to the band-edge states. However, the lower efficiency of the purely inorganic perovskites suggests that the presence of organic cation may be the key for achieving high PCE, although the microscopic details on how the organic cations enhance the efficiency is unclear at this stage.

\begin{figure}[t]
        \begin{center}
                \includegraphics[width=0.7\linewidth,clip]{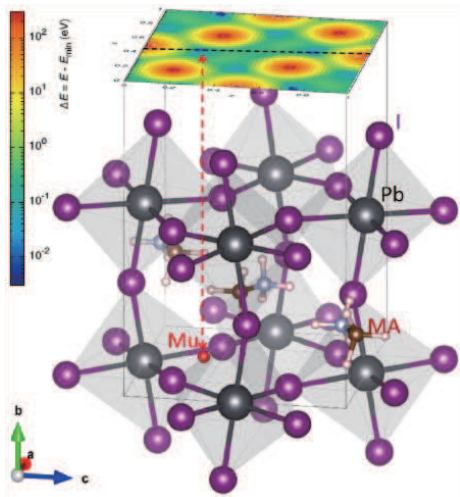}
\caption{The crystal structure of CH$_3$NH$_3$PbI$_3$ in the orthorhombic phase, where the organic cation (methyl ammonium, MA) is located in the center of a cage with PbI$_6$ octahedrons at the corners.  It is inferred from $\mu$SR and DFT calculations that the implanted muons occupy the (0.48,0.028,0.27) site in the unit cell (Mu, marked by a red ball). The color contour map shows the differential total formation energy for an interstitial hydrogen obtained by the DFT calculation.
}
\label{fig:structure}
\end{center}
\end{figure}

The crystal structure of HOIP is represented by that of MAPbI$_3$ consisting of 3D network of corner-shared PbI$_6$-octahedra and CH$_3$NH$_3^+$ molecule ions at the $A$ site in the generic perovskite structure $AB$O$_3$, where the orthorhombic structure ($Pnma$) observed at low temperatures is shown on Fig.~\ref{fig:structure}. There is now a consensus in the community that the long carrier lifetime is mainly due to the formation of large polarons\cite{Chen:16,Zhu:16}. The screened carriers are protected from scattering by defects and phonons, leading to the prolonged carrier lifetime \cite{Chen:16,Zhu:16,Chen:15,Chen:17}. the possibility of large polaron formation \cite{Chen:16,Zhu:16} by the reorientation of organic cations in response to the presence of photoexcited carriers \cite{Chen:15,Chen:17} is of particular interest in view of the cation molecular dynamics; the screened carriers are protected from scattering by defects and phonons, leading to the prolonged lifetime \cite{Chen:16,Zhu:16,Chen:17}. The microscopic mechanism of the screening must be associated with both lattice vibrations \cite{Chen:16,Zhu:16} and molecular rotations \cite{Chen:15,Chen:17} in response to the presence of photoexcited carriers. It is still an open question how much contribution each of the two make for the screening in HOIPs. Figuring out their contributions quantitatively is important because it will guide us in searching new materials for a better solar cell performance. It is not easy however to experimentally distinguish the two contributions. 

A fundamental difference between lattice vibrations and molecular rotations is related to their difference in coherence of dynamics; lattice vibrations are collective in nature with long spatial coherence while molecular rotations are mainly local and incoherent in space. Thus, a 
 local atomic probe that is sensitive to the molecular dynamics would be most useful, as it can provide information on molecules without relying on the coherence of their dynamics. 
Muon (Mu, a light isotope of H) spin relaxation technique does just that, 
as it probes the fluctuation of local magnetic fields at well-defined interstitial sites via muon spin rotation and relaxation ($\mu$SR) experiment. 
While the quasi-elastic neutron scattering \cite{Chen:15} and nuclear magnetic/quadrupole resonance (NMR/NQR, using $^2$D and $^{14}$N) \cite{Bernard:18} probe molecular motion over the time scale of 10$^{-12}$ s, 
$\mu$SR provides information over a unique time window of 10$^{-9}$--10$^{-5}$ s that may be relevant to the carrier dynamics. 
In this regard, the fact that NMR using $^{207}$Pb and $^{127}$I has not been successful so far due to fast relaxation rate ($1/T_2$) further justifies the application of $\mu$SR to probe the molecular motion using muons as a bystander.

Here, we report on the cation molecular dynamics in MAPbI$_3$ revealed by $\mu$SR measurements as a function of temperature from 60 K to 360 K.  
It is known that MAPbI$_3$ undergoes two structural transitions 
with increasing temperature \cite{Weller:15}, 
namely, the orthorhombic-to-tetragonal structural transition at $T_{\rm OT}\simeq162$~K 
which is followed by the tetragonal-to-cubic transition
at $T_{\rm TC}\simeq327$~K.  
We show that the $\mu$SR time spectra under zero/longitudinal field (ZF/LF) can be reasonably reproduced by the dynamical Gaussian Kubo-Toyabe relaxation function 
which describes the muon depolarization by the quasi-static linewidth $\Delta$ determined by the random local fields exerted from nuclear magnetic moments and its fluctuation rate $\nu$.   
Interestingly, a previous $\mu$SR study on MAPbI$_3$ reported that  the magnitude of $\Delta$ exhibits significant variation with temperature in response to the development of cation molecular motion above $\sim$100 K \cite{Ferdani:19}, whose microscopic details are yet to be understood.
Here, we provide a comprehensive account on how the local motion of cation molecules reduces the contribution of nuclear dipolar fields from $^1$H and/or $^{14}$N moments 
by the so-called motional narrowing effect.  In addition, we argue that a similar 
dynamical modulation is in effect for the electric dipole moments associated with MA molecules, and that the correlation between the relaxation time of MA molecular motion and the carrier lifetime infers the importance of low-frequency response in the local electric permittivity in the prolonged photo-induced carrier lifetime.

\begin{figure}[t]
        \begin{center}
                \includegraphics[width=0.75\linewidth,clip]{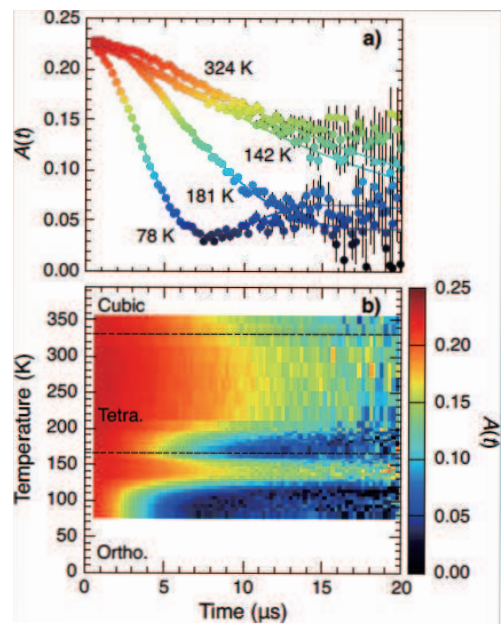}
                \caption{a) Typical examples of ZF-$\mu$SR time spectra ($\mu$-$e$ decay asymmetry) observed in MAPbI$_3$ at various temperatures, and b) the corresponding contour image of the asymmetry plotted on the time-temperature plane.}
\label{fig:MA-spectra}
\end{center}
\end{figure}

The conventional $\mu$SR measurements were carried out on two types of powder samples, 
one consisting of pristine MA,  and another partially substituted by deuterium (i.e., CD$_3$NH$_3^+$). The sample packed in aluminum foil was mounted on a silver sample holder which was attached to a He-flow cryostat for varying temperature over a range from 60~K through 360 K. The time evolution of the muon spin polarization was monitored by measuring the $\mu$-$e$ decay asymmetry, $A(t)$, using the ARTEMIS spectrometer furnished at the S1 area of J-PARC MUSE, Japan.  The DFT calculations were performed using the projector augmented wave approach \cite{Kresse:99} implemented in the Vienna ab initio simulation package (VASP) \cite{Kresse:96} with the Perdew-Burke-Ernzerhof (PBE) exchange correlation potential \cite{Perdew:96}, where the lattice parameters reported in the literature were adopted \cite{Weller:15}. The cutoff energy for the plane-wave basis set was  400 eV. The distribution of the
local magnetic field at the muon sites was calculated using Dipelec program \cite{Kojima:04}. The crystal structures were visualized using the VESTA program \cite{Momma:11}.

Figure \ref{fig:MA-spectra}a shows the ZF-$\mu$SR time spectra  [$A(t)$] observed at typical temperature points in MAPbI$_3$, which is complemented by Fig.~\ref{fig:MA-spectra}b showing the overall trend of the spectra versus temperature in a contour plot.
These spectra exhibit a slow Gaussian depolarization which is uniquely attributed to the random local fields from the nuclear magnetic moments. The initial asymmetry [$A_0=A(0)$] is close to that corresponding to $\sim$100\% muon polarization ($\simeq0.23$) irrespective of temperature, indicating that muons are mostly in the diamagnetic state (Mu$^+$ or Mu$^-$).
Considering that some of the incident muons stop at the backing material (silver) in which the depolarization is negligible, the time spectra are analyzed by curve-fits using the following function,
\begin{equation}
        A(t)=A_0 G_{\rm KT}(t;\Delta,\nu) e^{-\lambda t} + A_{\rm c},\label{kt}
\end{equation}
 which can be approximated for the case of $\nu\ll\Delta$ and a zero external field by
\begin{equation}
  A(t) \simeq A_0\left[\frac{1}{3}e^{-\nu t}+\frac{2}{3}(1-\Delta^2t^2)e^{-\frac{1}{2}\Delta^2t^2}\right]e^{-\lambda t}+A_{\rm c}.\label{kts}
\end{equation}
Here, $G_{\rm KT}(t;\Delta,\nu)$ represents the Gaussian Kubo-Toyabe relaxation function with $\Delta$ denoting the linewidth determined by the root mean square of the corresponding local field distribution, $\nu$ being the fluctuation rate of $\Delta$ \cite{Hayano:79}. The term
$e^{-\lambda t}$ is for the slow residual depolarization of unknown origin (which leads to a slight improvement of fits), and $A_{\rm c}$ is the background from the Ag sample holder. As shown in Fig.~\ref{fig:MA-spectra}a, $A(t)$ at 78 K exhibits the characteristic 1/3 term explicit in Eq.~[\ref{kts}], indicating that $\nu\ll\Delta$ at this temperature.

The linewidth $\Delta$ is determined by the sum of contributions from the $m$-th kind of nuclear magnetic moments ($m = 1,2,3$ and 4 for $^{1}$H, $^{14}$N, $^{127}$I, and $^{207}$Pb, whose natural abundance is nearly 100\%),
\begin{eqnarray}
\Delta_{\bm r}^2&\simeq&\gamma_\mu^2\sum_j\langle B_j^2\rangle 
=\gamma_\mu^2\sum_{j,m}\:\sum_{\alpha=x,y}\sum_{\beta=x,y,z}\gamma_m^2({\bm \hat{A}}_{j}{\bm I}_m)^2, \label{dlts}\\
{\bm \hat{A}}_{j}&=&A^{\alpha\beta}_{j}=(3\alpha_{j}\beta_{j}-\delta_{\alpha\beta}r_{j}^2)/r_{j}^5,\quad(\alpha, \beta=x,y,z)\nonumber
\end{eqnarray}
with $\gamma_\mu/2\pi=135.53$ [MHz/T] being the muon gyromagnetic ratio, ${\bm r}_{j}=(x_{j},y_{j},z_{j})$  the position vector of the $j$-th nucleus (with Mu at the origin), ${\bm \mu}_m=\gamma_m{\bm I}_m$ the nuclear magnetic moment with $\gamma_m$ being their gyromagnetic ratio. Because $^{14}$N and $^{127}$I nuclei have spin $I_m\ge1$, the corresponding ${\bm \mu}_m$ is subject to electric quadrupolar interaction with the electric field gradient generated by the point charge of the diamagnetic Mu. This leads to the reduction of effective ${\bm \mu}_m$ to the value parallel with ${\bm r}_{j}$ (by a factor $\sqrt{2/3}$ in the classical limit)~\cite{Hayano:79}.
We also conducted $\mu$SR measurements under a longitudinal field ($B_{\rm LF}$) up to 2~mT at each temperature point. The parameters in Eq.~[\ref{kt}] [common to the spectra with different $B_{\rm LF}$] were then determined reliably by simultaneous curve-fits of the spectra at various $B_{\rm LF}$.

\begin{figure}[t]
        \begin{center}
                \includegraphics[width=\linewidth,clip]{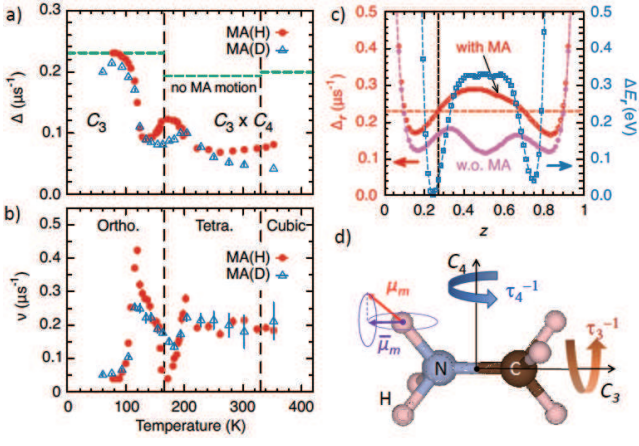}
                \caption{a) The Kubo-Toyabe linewidth ($\Delta$), and b) fluctuation rate ($\nu$) versus temperature, where circles and triangles represent data on samples consisting of CH$_3$NH$_3$ [MA(H)] and CD$_3$NH$_3$ [MA(D)], respectively.   c) The calculated linewidth ($\Delta_{\bm r}$) vs muon position (${\bm r}$) in the unit cell, where the $z$ axis is along the path connecting the potential minima shown by the dashed line in the contour plot of Fig.~\ref{fig:structure}.  The horizontal dashed line in c) represents experimental $\Delta$ corresponding to the static MA molecules (shown in Table~\ref{TBL:Delta}), with which the Mu sites are estimated (as marked by vertical lines).  The variation of the formation energy ($\varDelta E_{\bm r}$) is also plotted (right axis).  The horizontal lines in a) show $\Delta_{\bm r}$ at the most probable Mu sites without MA molecular motion (also in Table~\ref{TBL:Delta}). 
d) Schematic illustrations of MA molecule with two modes of jumping rotation, where the influence of rotation around the threefold/fourfold symmetry axis ($C_i$, $i=3,4$, with relaxation rate $\tau_i^{-1}$) is illustrated for nuclear magnetic moments ($\mu_m$). The contribution of $\mu_m$ to $\Delta$ is effectively reduced to $\overline{\mu}_m$ by motional averaging of the non-secular part around the $C_3$ axis, and then to zero by further averaging around $C_4$ axis when $\tau_i^{-1}\gg\Delta$ is satisfied with elevated temperature.   } 
\label{fig:MA-parameter}
\end{center}
\end{figure}

The temperature dependence of $\Delta$ deduced from the curve-fits using Eq.~[\ref{kt}] is shown 
in Fig.~\ref{fig:MA-parameter}a, 
where  $\Delta$ exhibits sharp decrease with increasing temperature above $\varTheta_3\simeq120$ K and $\varTheta_4\simeq190$ K across a broad hump around $T_{\rm OT}$.  The corresponding recoveries of asymmetry in the ZF-$\mu$SR spectra are also visible in Fig.~\ref{fig:MA-spectra}b.  Meanwhile, no significant change is observed around $T_{\rm TC}$.  It is often presumed that such behavior of $\Delta$ is due to the change of Mu sites; we note that $\lambda$ in Eq.~[\ref{kt}] remained to be small ($\le 0.02$ $\mu$s$^{-1}$) 
throughout the entire temperature range, supporting for the negligible ambiguity regarding the temperature dependence of $\Delta$ and $\nu$.
The Mu position (${\bm r}$) is then estimated by comparing $\Delta$ with $\Delta_{\bm r}$ calculated 
by Eq.~[\ref{dlts}] for the candidate sites suggested by the DFT calculations; the vicinity of the sites corresponding to the minima of the total formation energy ($E_{\bm r}$) for the Mu-MAPbI$_3$ system are examined for the respective structures. The problem associated with the small site occupancy of the MA molecules with varying orientation in the tetragonal and cubic structures are averted by substituting MA cation with Cs$^+$ which has a nearly equivalent ionic radius ($=0.18$ nm).

The linewidth $\Delta_{\bm r}$ and differential total energy $\varDelta E_{\bm r}$ [$=E_{\bm r}-E_{{\bm r}({\rm min})}$] calculated for three different phases with/without contribution of the MA molecules are shown in Fig.~\ref{fig:MA-parameter}c along the direction passing the $E_{{\bm r}({\rm min})}$ position near the Pb-I basal plane. 
In the orthorhombic phase, the Mu positions satisfying the condition $\Delta=\Delta_{\bm r}$ are close to that inferred from $E_{{\bm r}{\rm (min)}}$ which is located near the valley of the electrostatic potential surrounded by negatively charged iodines  (the corresponding contour plot for the orthorhombic phase is found in Fig.~\ref{fig:structure} top).  This indicates that Mu as pseudo-H is in a positively charged state (Mu$^+$).
The asymmetric tendency of $\varDelta E_{\bm r}$ along $x/z$ direction is attributed to the broken inversion symmetry of the MA molecules and associated local charge imbalance that leads to the electric dipole moment parallel to the three-fold rotational symmetry  ($C_3$) axis. 
It must be noted that $\Delta_{\bm r}$ is always greater than $\Delta$ over the region around $x/z=0.5$ at which $\varDelta E_{\bm r}$ exhibits local minima. Thus, we conclude that the Mu site in this phase is $(0.48,0.028,0.27)$ where $\Delta=\Delta_{\bm r}=0.230(1)$ $\mu$s$^{-1}$.
The estimation for the partially deuterated sample yields  $\Delta_{\bm r}=0.192$~$\mu$s$^{-1}$, where the reduction can be attributed to the smaller magnetic moment of $^2$H nuclei.

The Mu sites in the tetragonal ($I4mcm$) and cubic ($Pm\overline{3}m$) phases were also estimated by searching for the position satisfying the condition $\Delta=\Delta_{\bm r}$ near $\varDelta E_{{\bm r}({\rm min})}$. As a result, it turned out that $\Delta_{\bm r}$ was always greater than $\Delta$ when the contribution from the quasistatic MA molecules was included. This led us to conclude that $B_{\rm loc}$ from the MA cations was reduced by the motional averaging due to the jumping rotation of MA molecules themselves.  
According to the earlier studies using neutron scattering and NMR/NQR, the jumping rotation of MA molecules around the $C_3$ axis evolves for $\varTheta_3\le T\le T_{\rm OT}$ in the orthorhombic phase, which is followed by the onset of rotation around the four-fold symmetry ($C_4$) axis above $\varTheta_4$ \cite{Chen:15,Bernard:18}. 
The relaxation rate of these jumping rotations varies over a range much greater than $\Delta$ (i.e., $\tau_{3,4}^{-1}=10^{3}$--$10^{6}$ $\mu$s$^{-1}$) in the relevant temperature range, which is consistent with the fast fluctuation of $\Delta$. 
As is illustrated in Fig.~\ref{fig:MA-parameter}d, the rotation around the $C_3$ axis reduces the contribution of $\mu_m$ for $^1$H and $^{14}$N nuclei to $\overline{\mu}_m$ corresponding to the projection of $\mu_m$ to the $C_3$ axis when $\tau_3^{-1}\gg\Delta$; %
since the nuclear dipolar fields in Eq.~[\ref{dlts}] are expressed as
\begin{equation}
B_{j}=\frac{\mu_m}{r_{j}^3}[(3\cos^2\theta_{\bm r}-1)\cos\theta_{m}+
3\sin\theta_{\bm r}\cos\theta_{\bm r}\sin\theta_{m}\cos\phi_{m}],\label{hdip}
\end{equation}
where  $\theta_{\bm r}$ is the polar angle of ${\bm r}_{j}$, $\theta_{m}$ ($\phi_{m}$) are the polar (azimuth) angle of ${\bm \mu}_m$ measured from the $C_3$ axis, the term proportional to $\sin\theta_{m}$  in Eq.~[\ref{hdip}] is averaged out by the jumping rotation ($\langle\cos\phi_m\rangle\simeq0$) with remaining contribution $\overline{\mu}_m=\mu_m\cos\theta_{m}$ \cite{Hayano:79}. 
The contribution is eventually eliminated by further averaging around $C_4$ axis ($\perp C_3$) when $\tau_4^{-1}\gg\Delta$, leading to the reduction of the effective $\Delta$ in two steps. 
The small hump of $\Delta$ observed around $T_{\rm OT}$ can be interpreted as due to the Mu site change induced by the structural phase transition. 
This model allowed us to assign the most probable Mu site in the orthorhombic/cubic phases by the condition $\Delta=\Delta_{\bm r}$ with/without MA contribution for $\Delta_{\bm r}$.
For the tetragonal phase, we adopted the condition that $\Delta_{\bm r}$ without MA contribution was closest to $\Delta$. In Fig. 3a), $\Delta_{\bm r}$ for these Mu sites without MA molecular motion is shown for comparison. This allows us to clearly see the effect of MA molacular dynamics on $\Delta$.
 Table~\ref{TBL:Delta} summarizes $\Delta_{\bm r}$ for the candidate Mu sites for the each structural phase. 

\begin{table}[t]
\begin{tabular}{c|ccc}
\hline\hline
 \multirow{2}{*}{$A$ cation} & \multicolumn{3}{c}{$\Delta_{\bm r}$  ($\mu$s$^{-1}$)} \\
 & Orthorhombic  &  Tetragonal  & Cubic \\
 \hline 
CH$_3$NH$_3$ & 0.2302 & 0.1934 & 0.1989  \\
C--N &  0.1656 & 0.0948 & 0.0814\\
 null & 0.1455 &  0.0946  & 0.0812\\
CD$_3$NH$_3$ & 0.1920 &0.1542 & 0.1328 \\
\hline\hline
 \multirow{2}{*}{$\Delta$ (exp.)}  &0.2298(8) [78~K] & 0.1218(5) [171~K]  & 0.0813(9) [352~K] \\
& 0.091(1) [133 K]  &  0.0679(7) [260~K] & \\
\hline
Mu site & (0.48,0.028,0.27) &(0.54,0.044,0) & (0.58,0.58,0)\\
\hline\hline
\end{tabular}
\caption{The  Kubo-Toyabe linewidth ($\Delta_{\bm r}$) calculated for the most probable muon sites  in the respective structural phases of MAPbI$_3$ with a variety of $A$ cations. The lower rows are for $\Delta$ observed as extreme values (see Fig.~\ref{fig:MA-parameter}a) and atomic coordinates of the assigned Mu sites.}
\label{TBL:Delta}
\end{table}

Provided that the observed decrease in $\Delta$ above $\varTheta_3$ is mainly due to the jumping rotation of the MA molecules (around the $C_3$ axis), it is not the whole $\Delta$ that is fluctuating but the local field from the MA molecules. To understand the behavior of $\nu$ shown in Fig.~\ref{fig:MA-parameter}b in relation to the MA molecular motion  [NB: $\nu$ was obtained by curve-fit analysis using Eq.~[\ref{kt}] to allow its arbitrary variation], we phenomenologically extended the Kubo-Toyabe relaxation function to 
\begin{equation}
G_{\rm KT}(t;\Delta,\nu)\simeq G_{\rm KT}(t;\Delta_{\rm M},\nu_{\rm M})G_{\rm KT}(t;\Delta_{\rm T},\nu_{\rm T}),\label{kte}
\end{equation}
 where $\Delta_{\rm M}$ is the nuclear magnetic contributions from the MA molecules, $\Delta_{\rm T}$ [$=(\Delta^2-\Delta_{\rm M}^2)^{1/2}$] is the remaining quasistatic part, and $\nu_{\rm M}$ and $\nu_{\rm T}$ are the corresponding fluctuation frequencies of $\Delta_{\rm T}$ and $\Delta_{\rm M}$. Considering that $\nu_{\rm T}\ll\Delta_{\rm T}$ at low temperatures, we have
\begin{equation}
G_{\rm KT}(t;\Delta,\nu)\simeq\frac{1}{3}e^{-\nu_{\rm M}t}+\frac{2}{3}[1-\Delta_{\rm T}^2t^2+\frac{\Delta_{\rm M}^2}{\nu_{\rm M}}(e^{-\nu_{\rm M}t}-1)t]G_x(t),
\end{equation}
where $G_x(t)$ is the relaxation function under a transverse field \cite{Hayano:79,Kubo:54},
\begin{equation}
G_x(t)=e^{-\frac{1}{2}\Delta_{\rm T}^2t^2}\exp[-\frac{\Delta_{\rm M}^2}{\nu_{\rm M}^2}(e^{-\nu_{\rm M}t}-1+\nu_{\rm M}t)],
\end{equation}
which is exact for any $\nu_{\rm M}$. Then,
in the case of $\nu_{\rm M}\ll\Delta_{\rm T},\Delta_{\rm M}$, we have
\begin{equation}
G_{\rm KT}(t;\Delta,\nu)\simeq\frac{1}{3}e^{-\nu_{\rm M}t}+\frac{2}{3}[1-\Delta^2t^2]e^{-\frac{1}{2}\Delta^2t^2}.
\end{equation}
Therefore, the value of $\nu$ obtained from the fit by Eq.~[\ref{kt}] corresponds to $\nu_{\rm M}$. On the other hand, if $\nu_{\rm M}\gtrsim\Delta_{\rm M}$, Eq.~[\ref{kte}] can be roughly approximated to yield 
\begin{equation}
G_{\rm KT}(t;\Delta,\nu) \simeq G_{\rm KT}(t;\Delta_{\rm T},0)e^{-\frac{\Delta_{\rm M}^2}{\nu_{\rm M}}t},
\end{equation}
and the apparent decrease of $\nu$ obtained by fitting with Eq.~[\ref{kt}] can be attributed to the increase of $\nu_{\rm M}$, because $\nu$ as a fitting parameter is proportional to $\Delta_{\rm M}^2/\nu_{\rm M}$.
Thus, the behavior in $\nu$ seen for $\varTheta_3\le T\le T_{\rm OT}$ in Fig.~\ref{fig:MA-parameter}b can be interpreted as that due to the increase in the jump rotation frequency of the MA molecule around the $C_3$ axis. 

The change in $\nu$ for $\varTheta_4\le T\le 220$ K is also attributed to the similar mechanism acting on $\Delta_{\rm T}$ due to the increase of jumping rate around the $C_4$ axis, where $\Delta_{\rm M}$ and $\Delta_{\rm T}$  in Eq.~[\ref{kte}] are replaced by $\Delta_{\rm \overline{M}}$  (the remaining contribution from the MA molecules) and by $\Delta_{\rm L}$ [$=(\Delta_{\rm T}^2-\Delta_{\rm \overline{M}}^2)^{1/2}$, consisting only of the PbI$_3$ lattice contribution] with their fluctuation rate $\nu_{\rm \overline{M}}$ and $\nu_{\rm L}$, respectively. The increase of $\nu$ for $T_{\rm OT}\le T\le 200$ K is then understood as that of $\nu_{\rm \overline{M}}$, and the turnover above $\sim$200 K is described by $\nu\propto\Delta_{\rm \overline{M}}^2/\nu_{\rm \overline{M}}$.  The residual value of $\nu\simeq\nu_{\rm L}\simeq0.2$ $\mu$s$^{-1}$ at higher temperatures is attributed to the diffusion of iodine ions in the relevant temperature range\cite{Ferdani:19,Garcia:19}.

The variation of $\Delta$ in the tetragonal phase can be used to evaluate the reduction factor $x=\overline{\mu}_m/\mu_m$ by the motional effect using the relation 
\begin{equation}
\Delta_{\rm T}^2=\Delta_{\rm L}^2+x^2(\Delta-\Delta_{\rm L}^2).
\end{equation} 
Assuming that $\Delta=0.1934$ $\mu$s$^{-1}$ (calculated for the tetragonal phase), $\Delta_{\rm T}=0.1218(5)$ $\mu$s$^{-1}$ (at 171 K), and $\Delta_{\rm L}=0.0679(7)$ (at 260 K, which is close enough to $\Delta_{\bm r}=0.0946$ $\mu$s$^{-1}$ calculated without MA molecules), we have $x=0.56(3)$. 
Meanwhile, the amount of change in $\Delta$ between 78 K and $\sim$133 K [$=0.090(1)$ $\mu$s$^{-1}$] in the orthorhombic phase exceeds that predicted by quenching the contribution of MA molecules, which we tentatively attribute to the additional motion of Mu itself induced by the evolution of the relatively slow MA jumping rotation around the $C_{4}$ axis ($10^0\lesssim\nu\lesssim10^2$ $\mu$s$^{-1}$, not susceptible for neutron/NMR).  As shown in Fig.~\ref{fig:MA-parameter}c, the potential energy for Mu is asymmetric along the $C_{3}$ axis, and Mu tends to stay near the CH$_3$ bases. The fluctuation of the Mu potential induced by the MA reorientation around the $C_4$ axis will activate the Mu hopping between the equivalent sites in the unit cell, leading to the fluctuation of $\Delta_{\rm L}$. Note that this motion does not affect the averaging of $\Delta$ around the $C_3$ axis.

\begin{figure}[t]
        \begin{center}
                \includegraphics[width=\linewidth,clip]{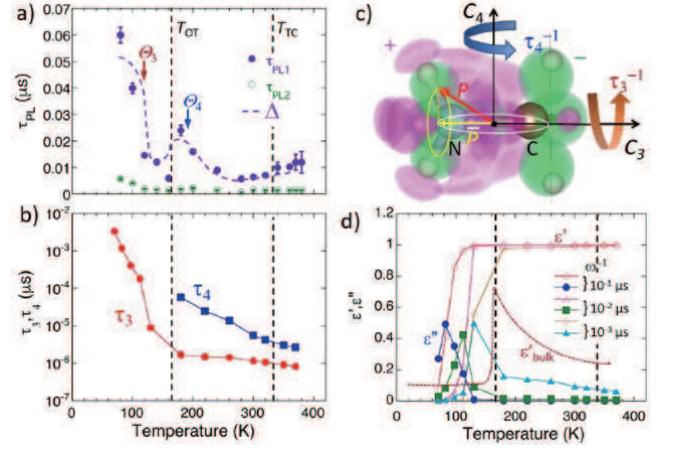}
                \caption{a) Photo-luminescence (PL) lifetime versus temperature in MAPbI$_3$ (quoted from Ref.\cite{Chen:17}), where the dashed curve indicates $\Delta$ ($^1$H) versus  temperature  shown in Fig.~\ref{fig:MA-parameter}a. b) The relaxation time $\tau_3$ ($\tau_4$) (in logarithmic scale) of the rotational motion around the $C_3$ ($C_4$) axis determined by quasi-elastic neutron scattering (after Ref.\cite{Chen:15}).  c) Schematic illustrations of MA molecule with two modes of jumping rotation, where the hatched areas indicate the local charge asymmetry obtained by the DFT calculation.  d) The contribution of MA molecules to the complex permittivity [$\varepsilon(\omega)=\varepsilon'+i\varepsilon''$] estimated by the Debye model (see text). The symbols in d) are predicted behavior of  $\varepsilon(\omega)$ when $\omega^{-1}$ is near the PL lifetime. $\varepsilon'_{\rm bulk}$ is the bulk permittivity (after Ref.\cite{Yamamuro:92}).}
\label{fig:PL-perm}
\end{center}
\end{figure}

It is remarkable that the overall temperature dependence of $\Delta$ including the hump around $T_{\rm OT}$ is in close resemblance with that observed for 
the photoluminescence lifetime \cite{Chen:17}.  As shown in Fig.~\ref{fig:PL-perm}a, the longer lifetime ($\tau_{\rm PL1}$) exhibits a sharp decrease above $\varTheta_3$ which is followed by a small hump around $T_{\rm OT}$ and the further reduction above $\varTheta_4$. A similar trend is observed for the shorter lifetime ($\tau_{\rm PL2}$).
Moreover, it is obvious in Fig.~\ref{fig:PL-perm}b that these behaviors are in parallel with the steep reduction of $\tau_3$ and subsequent onset and reduction of $\tau_4$ with increasing temperature.
Such correlations suggest an intrinsic relationship between the lifetime of photoexcited carriers 
and the MA molecular motion, which can be understood by considering that the mechanism causing the change in $\Delta$ is also in effect for the local dielectric permittivity $\varepsilon(\omega)$.
As is illustrated in Fig.~\ref{fig:PL-perm}c, the MA molecule has an electric dipole moment ${\bm P}$ along the $C_3$ axis due to the local charge imbalance, whose effective value seen from photoexited carriers is subject to reduction by the motional averaging on a certain time scale; it is reduced to $\overline{{\bm P}}$ (a projection to the C$_3$ axis) by jumping rotation around the $C_3$ axis, then to zero by additional rotation around $C_4$. 

Here, let us consider the contribution of the MA molecules using the Debye model \cite{Debye:29} in order to discuss the relationship with the dielectric permittivity in more detail. The dielectric response of  non-interacting dipoles is described by the complex permittivity $\varepsilon(\omega)=\varepsilon'(\omega)+i \varepsilon''(\omega)$ with 
\begin{eqnarray}
\varepsilon'(\omega)&=&\varepsilon_\infty+(\varepsilon_s-\varepsilon_\infty)\frac{1}{1+\omega^2\tau^2},\\
\varepsilon''(\omega)&=&(\varepsilon_s-\varepsilon_\infty)\frac{\omega\tau}{1+\omega^2\tau^2},
\end{eqnarray}
where $\varepsilon_s$ $[=\varepsilon(0)]$ is the static permittivity, $\varepsilon_\infty=\varepsilon(\infty)$, and $\tau$ is the relaxation time of the electric dipoles.  We assume that $\tau$ for the MA molecules is determined by the mean value $\tau^{-1}_{\rm MA}=\tau_3^{-1}+\tau_4^{-1}$. While $\varepsilon_s-\varepsilon_\infty\simeq N|{\bm P}|^2/3k_BT$ for the free dipoles (with $N$ being the number of dipoles in the unit volume), we presume that the temperature dependence for the MA molecules is represented by that of $\tau_{\rm MA}$ via $\tau_i$ shown in Fig.~\ref{fig:PL-perm}b.  Assuming that $\varepsilon_\infty=0$, the calculated $\varepsilon'$ and $\varepsilon''$ (normalized by $\varepsilon_s$) versus temperature is shown in Fig.~\ref{fig:PL-perm}d for a variety of $\omega^{-1}$ relevant with $\tau_{{\rm PL}i}$.   The coincidence between $\varTheta_3$ and the temperature where $\varepsilon''(\omega)$ exhibits a peak observed for $\omega\sim10^2$ $\mu$s$^{-1}$ suggests that the inelastic (energy exchanging) interaction between the photoinduced carriers and MA cations in this frequency range is a crucial factor in determining the carrier lifetime. Such a low frequency response is expected to help reorienting MA molecules in response to the Coulomb interaction with carriers, serving as an electric screening due to the local permittivity.  Meanwhile, the static component ($\varepsilon'$) shows the least dependence on temperature [$\varepsilon'(\omega)\simeq\varepsilon_s$ for $\omega\le10^3$ $\mu$s$^{-1}$], contributing to the bulk static permittivity ($\varepsilon'_{\rm bulk}$) as a constant offset.  The Curie-Weiss behavior of $\varepsilon'_{\rm bulk}$ reported in the literature \cite{Yamamuro:92} is then attributed to the translational displacement of MA molecules against the Pb-I lattice, where the displacement is unlocked by the onset of the fast jumping rotation around the $C_4$ axis. 

Finally, as inferred from the temperature dependence of $\Delta$ and $\tau_3$, the characteristic temperature ($\varTheta_3$)  where the MA molecular motion exhibits sharp enhancement is significantly lower than $T_{\rm OT}$. Within the above scenario, this  suggests that the structural phase transition is driven by the MA molecular motion \cite{Chen_ScieceAdv}, which is in line with the shift of the hump in $\Delta$ to a higher temperature ($\sim$200 K) for the deuterated MA in which a higher $T_{\rm OT}$ is also inferred from the previous diffraction study \cite{Whitfield:16}.  The similar situation is then speculated for the tetragonal-to-cubic transition, which may be important to consider the relative stability of the PbI$_3$ frame structure.

In conclusion, our detailed $\mu$SR study on MAPbI$_3$ that is a local probe as a function of $T$ clearly shows that the molecular rotations make the major contribution to the formation of large polarons and thus to the long carrier lifetime in the HOIP.


\begin{acknowledgments}
We would like to thank the MLF staff for their technical support. Thanks are also to Senku Tanaka for fruitful discussion during data analysis and to Hua Li for the DFT calculations. This work was supported by the MEXT Elements Strategy Initiative to Form Core Research Centers, from the Ministry of Education, Culture, Sports, Science, and Technology of Japan (MEXT) under Grant No. JPMXP0112101001. M.H. also acknowledges the support of JSPS KAKENHI Grant No.19K15033 from MEXT.  The $\mu$SR experiments were conducted at the Materials and Life Science Experimental Facility (MLF), J-PARC under the support of Inter-University-Research Programs (Proposals Nos. 2017MI21, 2018B0075) by Institute of Materials Structure Science, KEK. S.-H. L. and J. J. C. acknowledge support from the U.S. Department of Energy, Office of Science, Office of Basic Energy Sciences under Award Number DE-SC0016144.
\end{acknowledgments}

%

\end{document}